# Round-the-Clock Sub-Ambient Cooling via Dynamic Sky View Factor Steering

**Qiuyu Chen[1,#], Minghao Dong[1,#,*], Zheng Zhang[1], Xiaodong Zhao[1], Peng Xiao[2,*] and Zhen Chen[1,*]**

1. Jiangsu Key Laboratory for Design and Manufacturing of Precision Medicine Equipment, School of Mechanical Engineering, Southeast University, Nanjing 211189, China
2. State Grid Electric Power Research Institute, State Grid Jiangsu Electric Power Co., Ltd., Nanjing, 211103, China

**# These authors contributed equally to this work**

*** Leading contact:** zhenchen@seu.edu.cn (Z.C.)

**Co-corresponding:** mhdong@foxmail.com (M.D) and vodoco@foxmail.com (P.X.)

# Abstract

Passive daytime radiative cooling holds significant potential to address sustainability challenges such as the energy-water nexus. However, common static horizontal configurations cannot respond to dynamic changes of environmental conditions such as solar position and cloud coverage, thereby forfeiting the opportunity of optimizing the cooling performance in the 24-h day-night cycle. Here, inspired from the heliotropism of sunflowers, we develop a dynamic sky view factor steering (DSVFS) system. By dynamically steering the emitter to an optimal angle, this system not only relaxes the stringent dual requirement of ultra-high solar reflectivity and infrared emissivity of the radiative emitter, but also maximizes its cooling power round-the clock. Using this DSVFS system, we experimentally demonstrate sub-ambient cooling during a typical hot noon even with a near-blackbody emitter; with a selective emitter, we demonstrate an increase of cooling power by 135% as compared to its static counterpart. Case studies of multiple cities across the world indicate a maximum annual electricity saving of up to 200 kWh/m$^2$.

# Introduction

In the context of contemporary energy and environmental crises, fully passive radiative cooling technology—consuming zero energy and emitting no greenhouse gases—has recently garnered significant attention for its potential in thermal management of buildings,[1–4] personal thermal comfort,[5,6] water harvesting[7–11] and food production.[12,13] This technology exploits the ultracold outer space at approximately 3 Kelvin and the fact that the peak of the blackbody emission at or near 300 kelvin lies in the atmospheric transparency window in the wavelength range between 8 to 13 μm.[14–19].

Radiative emitters on horizontal surfaces maximize the sky view factor, thus enhancing the net emission of infrared thermal radiation to outer space. However, incoming solar irradiance significantly compromises, if not entirely prevents, sub-ambient cooling performance. Consequently, sub-ambient daytime radiative cooling necessitates that the solar reflectivity of horizontal emitters surpasses 90%, even with optimal selective mid infrared (MIR) emission.[20] This dual requirement of solar reflectivity and MIR emissivity imposes stringent demands on photonic design, leading to a significant challenge—and often a practical hurdle—for real-world implementation.

To relax this stringent requirement, tilting or shading the radiative emitter, which avoids direct sunlight and thus significantly reduces the demand for ultra-high solar reflectivity, presents an effective strategy, although this strategy sacrifices a portion of the sky view factor.[21–28] For example, Chen *et al*. applied an infrared-transparent solar absorber on top of an infrared selective emitter to simultaneously and synergistically harvest energy from the sun and outer space.[25] Pei group achieved sub-ambient cooling using emitters that are not optimally engineered by exploiting declined surfaces to avoid direct sunlight exposure, and subsequently obtained an optimized angle by summarizing their measurements.[26,27] Yet static tilts cannot adapt to dynamic change of environmental conditions such as the solar position and the cloud coverage. To address this shortcoming, Bhatia et al. enhanced daytime radiative cooling by employing a dynamic shade that adjusts its position with the sun to block direct solar irradiation.[28] However, large-scale implementation of this design requires extensive shading structures, which severely degrade the sky view factor at the central emitter region, weakening its radiative cooling performance and ultimately preventing sub-ambient cooling (figure S8).

To overcome this limitation, we develop a bio-inspired system for dynamic sky view factor steering (DSVFS) system. This system continuously aligns the emitter toward the optimal orientation to maximize the cooling power. Using this DSVFS system, we achieve round-the-clock sub-ambient radiative cooling even with a near-blackbody emitter, demonstrating a maximum temperature reduction of 9.5℃ below ambient and a mean cooling power of $57.2 \mathrm{W/m^2}$ during a typical hot noon. Next, equipping a well-designed spectral selective emitter with this DSVFS system, we experimentally achieve a cooling power increase by $40.8 \mathrm{W/m^2}$, as compared to its horizontal counterpart. Finally, case studies across multiple cities in both the USA and China show that our system can yield summer energy savings up to 200 $\mathrm{kWh/m^2}$, underscoring its significant energy-saving potential.

# Results

Conventional radiative cooling devices typically maintain static horizontal orientations. This static design involves two major assumptions: (i) near-ideal selective spectral properties in materials, and (ii) the unobstructed heat transfer between the emitter and outer space, enabling maximum cooling power capture through full hemispherical sky access. However, when implementing either imperfect radiative properties of emitters or partial obstruction in sky access, active field-of-view (FOV) optimization becomes essential for continuous 24-hour high cooling performance. Here we present a setup that can actively maximize its net cooling power under real-time sky exposure through dynamic sky view factor steering (DSVFS).

**Design of Dynamic Sky View Factor Steering (DSVFS) System**

The system (figure 1) dynamically adjusts according to four parameters: the solar irradiation, the atmospheric radiation, the environmental thermal radiation, and deployment scenarios. At clear-sky daytime, the optimal radiative cooling field- of-view (FOV) critically depends on spectral properties of emitters. Theoretical analysis confirms that for emitters with solar absorptivity >0.04 and mean emissivity of 0.95 (4–20 μm), optimal orientation requires maintaining parallel alignment with the direct sunlight to minimize solar absorption (Supplemental Information, Section 6). Under overcast sky or at night, the dynamic sky view factor steering (DSVFS) system returns to the horizontal position to maximize the net radiative cooling power, as the solar irradiation becomes isotropic or vanishes.

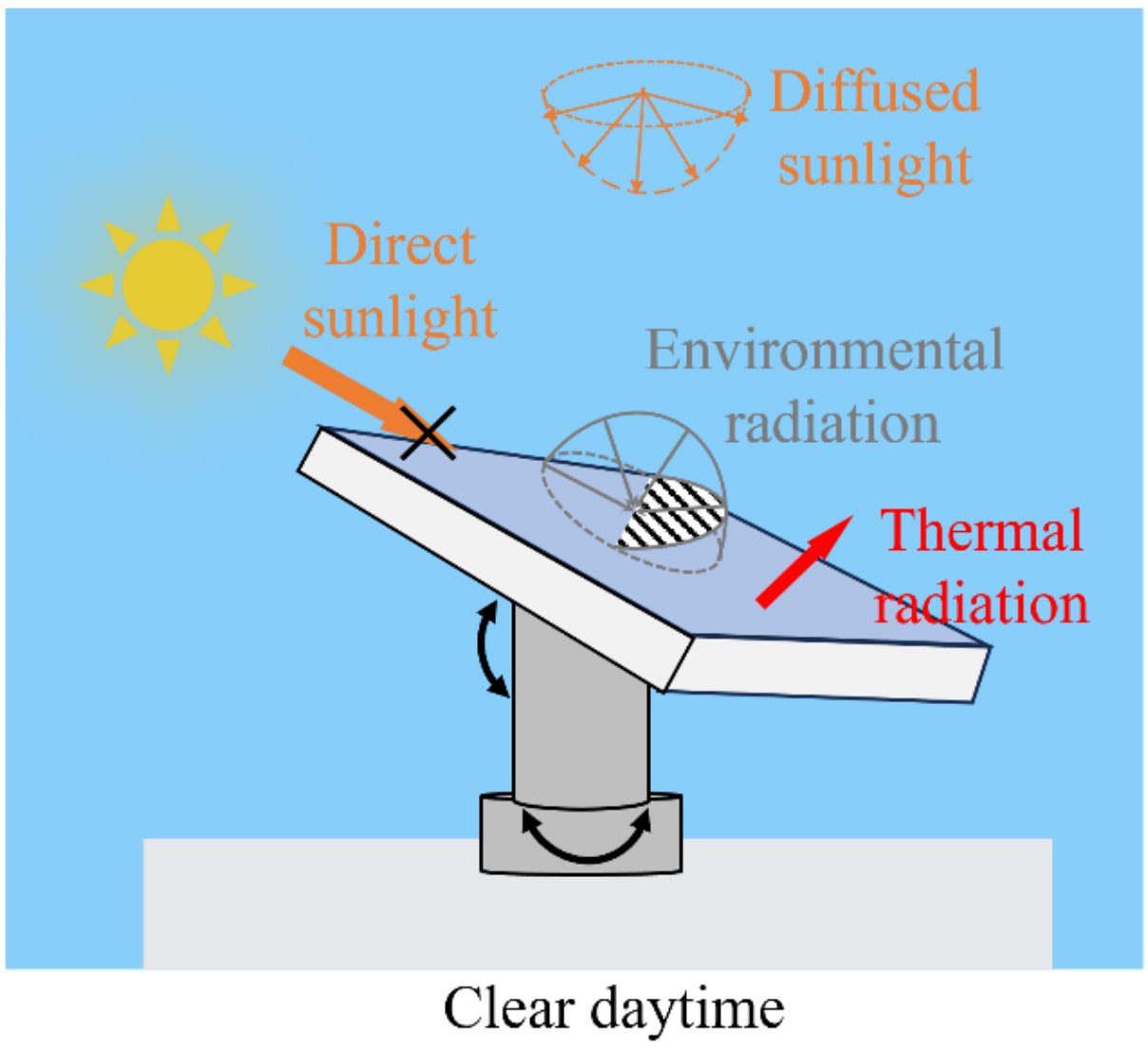

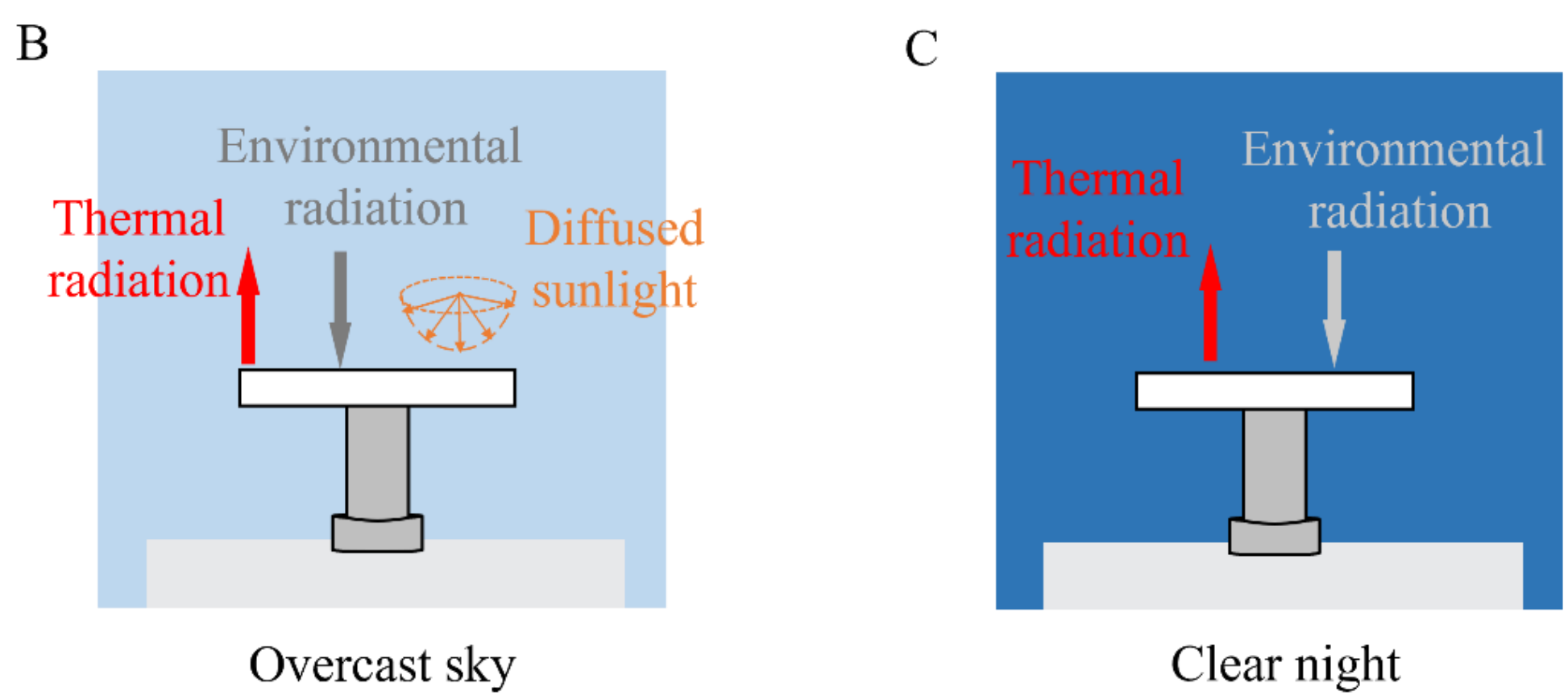

**Figure 1. Concept of Dynamic Sky View Factor Steering (DSVFS).**
**(A)** At clear daytime, the system orients itself to avoid the direct sunlight, minimizing solar absorption. **(B)** Under overcast sky and **(C)** At clear night, the system aligns horizontally, maximizing the radiative heat exchange with the cold outer space.

As shown in figures 2A-B, we construct the DSVFS system under representative rooftop scenarios. To demonstrate the advantages of this DSVFS configuration, a near-blackbody emitter (prepared with commercial black paint), which is commonly accepted to be unable to achieve sub-ambient daytime cooling, is analyzed. The DSVFS system maintains coplanar alignment of the emitter surface with direct solar irradiance vectors through continuous two-axis motor adjustments driven by real-time solar coordinates from integrated sensors during daylight. This dynamic adjustment persists until achieving the geometric condition where the declined angle $(\beta)$ complements the solar zenith angle $(\theta_z)$ with $\beta = 90° - \theta_z$ and the declination axis center aligns directly toward the sun, thereby minimizing solar incidence through optimal angular positioning. At night, without sunlight arriving at the sensor, the device returns to the original horizontal position. The real-time declined angle is measured using an angle sensor fixed on the plane of the emitter. A dynamometer is installed to measure the operating power consumption of the device. Given that the diffuse solar radiation (≥16.5% of total irradiance even under clear skies[29]) remains unavoidably absorbable by the DSVFS system, a 25-μm-thick nano-porous polyethylene (nanoPE) optical barrier encapsulates the emitter, suppressing the residual diffuse solar irradiation while geometrically confining a 5 cm air gap to minimize conductive heat transfer (figure S1; Supplemental Information, Section 1). Polystyrene foams are employed both on the bottom and lateral surfaces to minimize the non-radiative heat loss. All the other surfaces of the insulation device exposed to the air are covered with aluminized Mylar to minimize radiation heat transfer with the environment.

As DSVFS system dramatically reduces the daytime solar irradiation through solar projection control, the black paint exhibits an enhanced radiative cooling flux exceeding 700 W/m² at solar noon (figure 2C; see supplemental Information, Section 7). The cooling flux progressively increases as the solar descends until attaining parity with the horizontal emitter configuration post-sunset, concurrent with DSVFS system reverting to the baseline horizontal alignment. The system demonstrates a daily accumulated cooling energy of 1 kWh/day (figure 2D; Supplemental Information, Section 7), showing an improvement of 140% relative to the horizontal reference configuration. During summer solstice periods under peak insolation conditions (>1000 W/m²), the radiative cooling system demonstrates maximal diurnal energy saving, with DSVFS enhancements approaching 3 kWh/day.

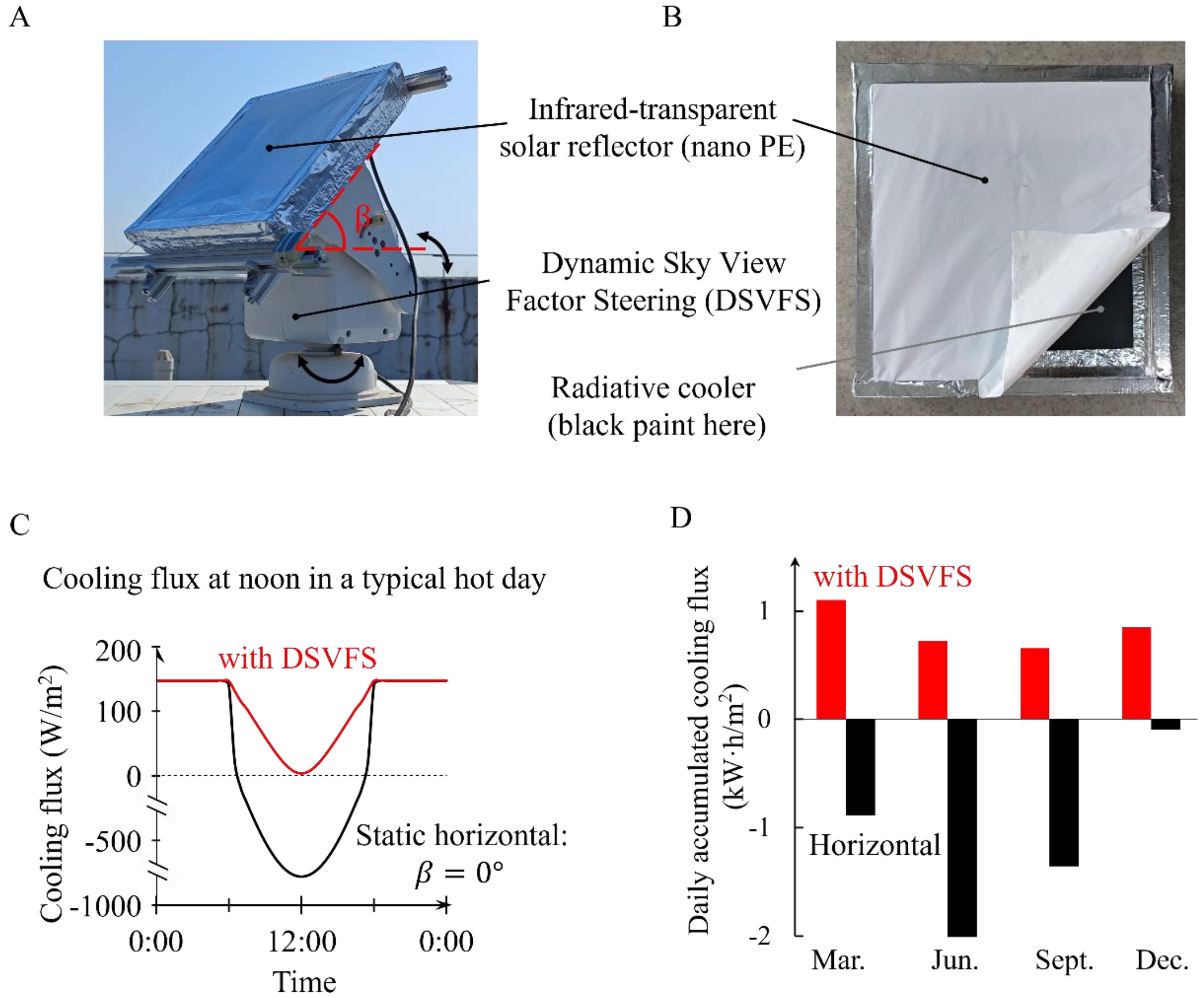

**Figure 2. Experimental Setup and Predicted Cooling Performance of a Near-blackbody Emitter**

**equipped with DSVFS system.**

**(A-B)** Experimental apparatus, integrating a near-blackbody emitter (black paint), an infrared-transparent solar reflector (25-µm-thick nanoporous polyethylene film, nano PE), and the DSVFS module that comprises a solar sensor and a dual-axis gimbal to enable self-adaptive, anti-heliotropic steering.

**(C)** Calculated net cooling flux of the near-blackbody emitter equipped with the DSVFS (red), with the static horizontal setup (black) for reference. Calculations assumes realistic spectra (see figure 3a) on a typical hot day (ambient temperature, $T_{ambient}$ = 30°C and relative humidity, RH = 30%).

**(D)** Comparison of the daily accumulated cooling flux in a 24-h day-night cycle across four representative months. These calculations highlight the significant potential of achieving sub-ambient radiative cooling even with emitters that are not optimally designed.

## Experimental Results.

The outdoor experiments are conducted in Nanjing, China, with two representative emitters: a near-blackbody emitter made by an aluminum plate coated with commercial black paint), and a selective emitter made by an aluminum plate coated with home-made radiative cooling paint (RC paint) [30]. Figure 3 shows the spectra of the transmission of nano PE, the emissivity of the near-body emitter and the selective emitter. The weighted average solar absorptivity and 8-13 µm emissivity of the black paint is 99.6% and 98.5%. That of the RC paint is respectively 4.6% and 98.5%. The nano PE functions as a reflector of the diffuse-solar irradiation with a weighted average solar reflectivity of 51.4% in the solar range and a weighted average transmittance of 94.9% in the atmospheric transparent window (8-13 µm), which allows IR-emission of the emitter to penetrate. Two devices, one equipped with the DSVFS system and the other fixed in the horizontal position, are placed sufficiently far apart to prevent interactions.

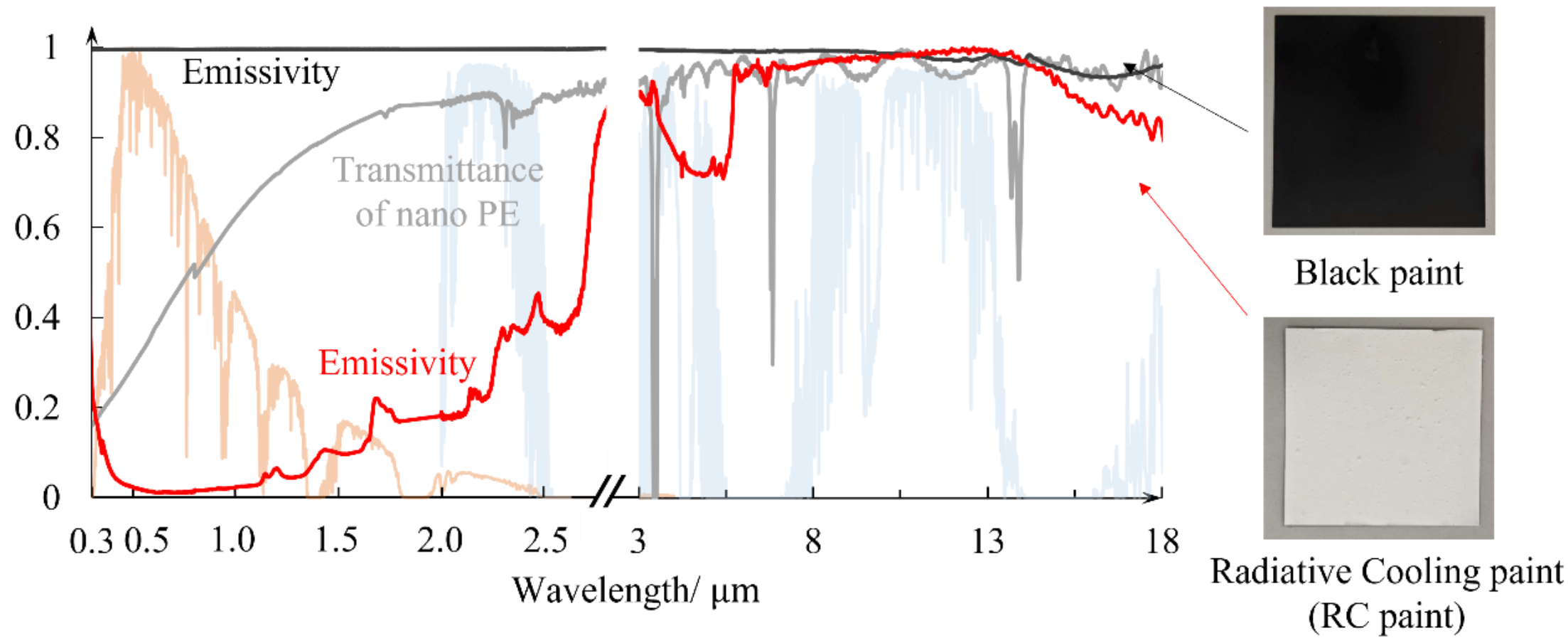

**Figure 3. Spectra of key ingredients.**

Emissivity of the black paint (black) and the homemade radiative cooling paint (red), and the transmittance of the nano PE (grey), with the normalized solar irradiance (orange) and the atmospheric transmittance (light blue) as references.

Figure 4A shows the temperature and the cooling flux of the black paint during a continuous day- and- night measurement. As expected, due to the high solar absorption, the near-blackbody emitter fails to achieve sub-ambient cooling at the fixed horizontal position (black lines in figure 4A). In contrast, the emitter with DSVFS system maintains a continuous average cooling of 13.6 ℃ below the horizontal emitter and 6.2 ℃ below the ambient temperature throughout the daytime. At noon, when solar intensity peaks, the emitter with DSVFS system reaches a temperature of 23.9 ℃ lower than the horizontal emitter and 5.1 ℃ below the ambient, indicating that this setup greatly enhances the cooling performance of the emitters. As solar irradiance intensifies, the temperature difference between emitters peaks at noon due to increased solar avoidance enabled by the DSVFS system. However, the temperature difference between the emitter with DSVFS system and the ambient temperature reaches its minimum at noon. On one hand, diffuse solar radiation peaks at noon, which raises the temperature of the emitter with DSVFS system. On the other hand, the declined angle increases as the sun rises. As the declined angle increases, the view factor towards the sky shrinks; the infrared thermal emission through the sky decreases. During night, the temperature of two emitters remains almost the same and the mean temperature difference is less than 0.1 ℃, confirming the same heat transfer structure and emitter spectra of two devices in both experiments (Supplemental Information, Section 2). This confirms that the temperature difference between two devices at daytime is due to the DSVFS system.

The real-time cooling flux is also measured. A heating patch (800 × 800 mm) is attached between the emitter and polystyrene foams to heat the emitter to the ambient temperature, which is controlled by a Proportional-Integral-Derivative (PID) system (Supplemental Information, Section 2). According to equation (1), the real-time cooling flux of the emitter equals to the input heating power of the heater. Figure 4C presents the measured cooling flux of the emitter with DSVFS system (red line). The horizontal emitter with black paint exhibits no sub-ambient cooling flux at daytime, while calculated cooling flux of the horizontal emitter (dashed black line) is shown as reference. For the black paint test, the mean temperature difference between the emitter and the ambient is 0.1 ℃, confirming the reliability of the measured cooling flux. The emitter with DSVFS system achieves a radiative cooling flux of 36.9 $\mathrm{W/m^2}$ at noon. In contrast, the calculated cooling flux of the horizontal emitter is -321.8 $\mathrm{W/m^2}$. The cooling flux of the emitter with DSVFS system varies against the solar intensity, reaching a minimum at the peak of the solar intensity, as the absorbed diffuse solar energy increases and the environmental radiation rises. However, the variation of the cooling flux is less pronounced as compared to the solar intensity, as the direct solar radiation is mostly avoided and the diffuse solar radiation dominated.

To examine the cooling enhancement of the selective emitter using DSVFS system, a similar experiment is also performed using homemade radiative cooling paint, which is exposed directly to the atmosphere without heat insulation (Supplemental Information, Section 1). Figures 4B and 4D show the experimental results of the radiative cooling paint. The radiative cooling paint can achieve sub-ambient cooling (figure 4B), regardless of whether DSVFS system is used or not. The radiative cooling paint with DSVFS system maintains a continuous average cooling of 0.8 °C below the horizontal paint and 3 °C below the ambient temperature throughout the daytime. At noon, when solar irradiance peaks, the emitter with DSVFS system reaches a temperature of 1.1 °C lower than the horizontal emitter and 2.2 °C below the ambient, indicating that the DSVFS system further improves the cooling ability of high-performance radiative cooling materials. As shown in figure 4D, the selective emitter with DSVFS system achieves 71 $W/m^2$ at solar noon, while the horizontal emitter reaches 30 $W/m^2$, quantifying a cooling flux amplification of 135% using DSVFS system. We note that although the DSVFS system consumes an average power of 2 $W/m^2$, the cooling flux enhancement ranges from 40 to 360 $W/m^2$ for different emitter types, indicating a great potential of the DSVFS mechanism. We also perform an experiment on a cloudy day using black paint to demonstrate the cooling performance of the DSVFS system under different weather conditions (see figure S10).

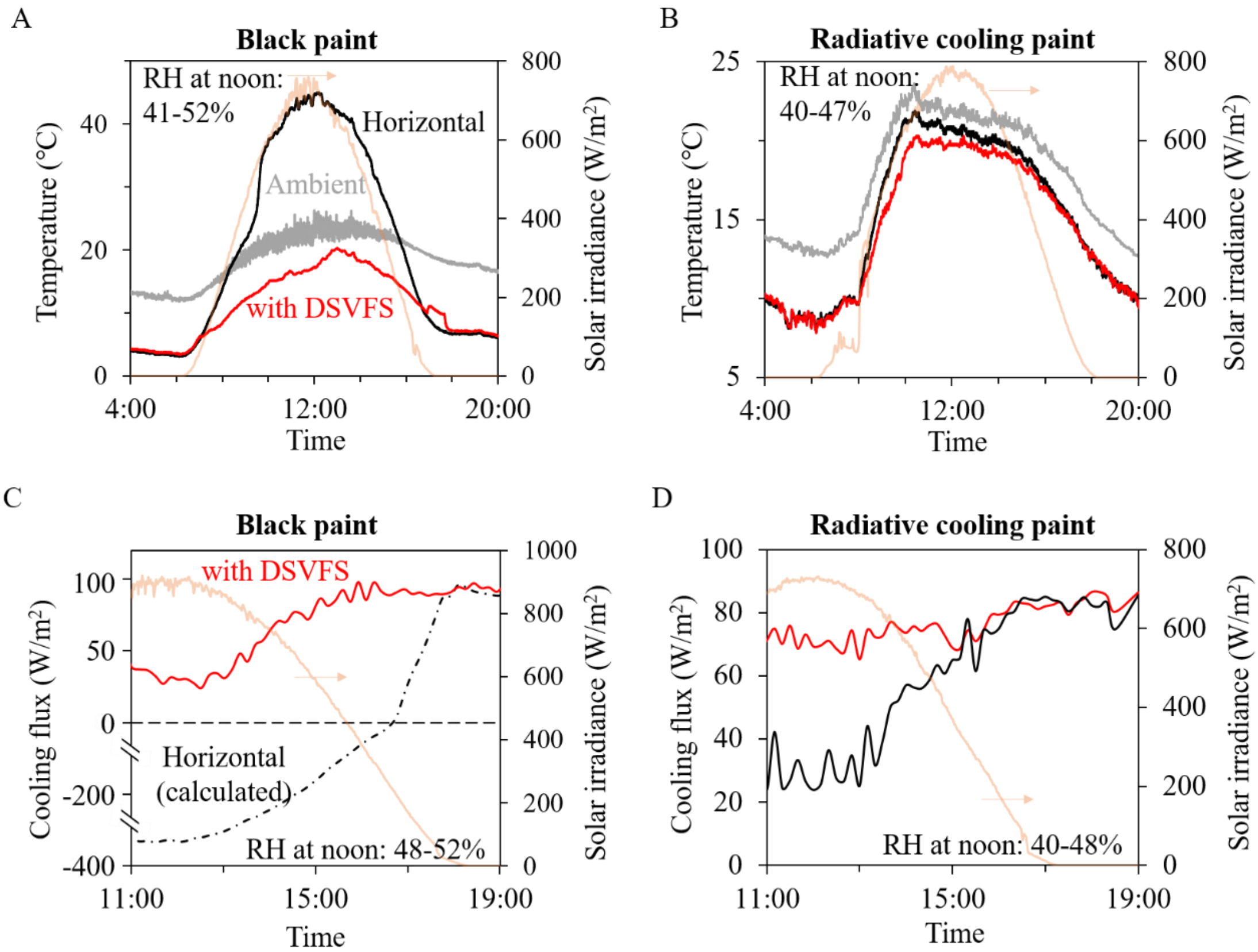

**Figure 4. Experimental Results with DSVFS system.**

**(A)** Temperature and **(C)** Cooling flux of the black paint: fixed horizontally (black line) vs. equipped with the DSVFS system (red line).

**(B)** Temperature and **(D)** Cooling flux of the radiative cooling paint: fixed horizontally (black line) vs. equipped with the DSVFS system (red line). The ambient temperature (gray lines) and solar irradiance (orange lines; right axis) are provided as references.

## Methods.

Here, we develop a heat transfer model for declined surfaces to further verify the energy saving potential of the dynamic sky view factor steering (DSVFS) configuration.

When the declined angle of the emitter is equal to or greater than the elevation angle ($\beta = \eta = 90° - \theta_z$), the direct solar irradiation would not be absorbed by the emitter. However, with the increase of declined angle, the view factor between the emitter and the sky decreases, leading to a reduction of the thermal emission power through the transparent window. A complete theoretical model is established here to investigate the daytime radiative cooling effects of declined emitters with different spectra.

The daytime radiative cooling flux $P_{cooling}(T, \beta)$ of the emitter at temperature $T$ can be expressed by considering all contributions to the energy balance:

$$P_{cooling}(T, \beta) = P_{rad}(T) - P_{atm}(T_{amb}, \beta) - P_{ground}(T_g, \beta) - P_{solar}(\beta) - P_{para}(T, T_{amb}) \quad (1)$$

Here, $P_{rad}$ denotes the radiated power radiated by the emitter, $P_{atm}$ denotes the absorbed atmospheric radiation by the emitter, $P_{ground}$ denotes the absorbed ground radiation by the emitter, and $P_{solar}$ denotes the absorbed solar irradiation by the emitter. The thermal emission from the declined emitter above the horizon exchanges heat with the atmosphere while that below the horizon exchanges heat with the ground. $P_{para}$ denotes the parasitic heat exchange with the environment.

The calculation of the atmospheric radiation is complex because the declination breaks the symmetry of radiation. Establishing the angular dependence of directional emissivity of the emitter and the anisotropic atmospheric radiance is required. In the coordinate system aligned with the declined surface, the zenith angle of the atmospheric radiance is denoted $\theta$, differing from $\gamma$ in the original coordinate system (Supplemental Information, Section 4). The atmospheric radiance absorbed by the emitter can be evaluated as shown in Equation 2.

$$p_{atm}(T_{amb}, \beta) = \int_0^{\pi} d\varphi \int_0^{\theta(\varphi)} I_{atm}(T_{amb}, \lambda, \gamma, H)\varepsilon_{emitter}(\lambda, \theta) \sin\theta \cos\theta \, d\theta d\lambda + \int_{\pi}^{2\pi} d\varphi \int_0^{\frac{\pi}{2}} I_{atm}(T_{amb}, \lambda, \gamma, H)\varepsilon_{emitter}(\lambda, \theta) \sin\theta \cos\theta \, d\theta d\lambda \quad (2)$$

where $\theta(\varphi)$ denotes the function of the declined surface ($\sin^2\theta = \frac{1}{1+\sin^2\varphi \tan^2\beta}$). $T_{amb}$ denotes the ambient temperature. $I_{atm}$ represents the directional atmospheric radiance, which could be calculated by Modtran[31]. H denotes the altitude.

The ground radiance absorbed by the emitter is assumed isotropic and can be evaluated as shown in Equation 3.

$$p_{ground}(T_g, \beta) = \int_0^{\pi} d\varphi \int_{\theta(\varphi)}^{\frac{\pi}{2}} I_{BB}(T_g, \lambda)\varepsilon_{ground}\varepsilon_{emitter}(\lambda, \theta) \sin\theta \cos\theta \, d\theta d\lambda \quad (3)$$

where $I_{BB}$ denotes the spectral radiance of a blackbody; $T_g$ denotes the temperature of the ground. $\varepsilon_{ground}$ denotes the emissivity of the ground, here 0.5 is taken[29]; $\varepsilon_{emitter}$ denotes the spectral directional emissivity of the emitter.

Using this model, we find that the DSVFS system performance exhibits strong dependence on two primary factors, the emitter's spectrum and the diurnal solar zenith angle variations. Emitters with higher solar absorptivity exhibit greater cooling flux enhancement through declination optimization, achieving greater solar energy reduction compared to low-absorptivity emitters because more solar energy is avoided.

The critical threshold of solar absorption is 0.02. When the solar absorption is lower than this value, the negligible absorption in horizontal orientations renders the declination angle adjustments ineffective (Supplemental Information, Section 6). At smaller solar zenith angles (sun higher in the sky), the surface-level solar radiation intensifies, causing the emitter to absorb more solar energy. Consequently, the cooling flux enhancement achieved by declination is larger than that the solar zenith angle is small. Analysis also shows that the optimal declined angle is complementary to the solar zenith angle for most emitters unless the solar reflectivity of the emitter is extremely high (Supplemental Information, Section 6).

**Potential of Large-Scale Implementation**

Our experimental demonstration provides a simple and low-cost approach to enhance the performance of daytime radiative cooling, which can be implemented in a large area. This method essentially minimizes the solar projected area on the emitter, thus avoiding the absorption of the dominant direct solar irradiation. Unlike the traditional method of optimizing the spectrum of the emitter that focuses on maximizing the solar reflectivity, this setup enables commercially available materials to realize sub-ambient daytime radiative cooling and enhances the cooling performance of radiative cooling materials.

To verify the universal applicability of the DSVFS system across diverse geographical regions, we calculate the summer radiative cooling energy savings of 31 Chinese cities and 51 USA cities from June to September (figure 5; see section 7 of the Supplemental Information for details). These calculations assume an emitter covered with a commercial-grade paint (with solar reflectivity of 75.7%; see figure S9). Most areas in both countries could save more than 150 $\mathrm{kWh/m^2}$ in summer. In both countries, the energy savings are greater in the western region than that in the eastern region, owing to the former's higher altitude and greater solar energy availability, which collectively enhance the cooling energy reduction. Figure 5B shows the theoretical analysis of the summer cooling performance of a horizontally oriented ideal emitters (blue dotted) with a solar reflectivity of $\rho$=1 between 0 and 4$\mu$m and an infrared emissivity of $\varepsilon$=1 elsewhere, the commercial paint aligned horizontally (black) vs. equipped with the DSVFS system (red), of four representative Chinese and USA cities. The commercial paint with the static horizontal orientation cannot achieve sub-ambient cooling, leading to a negative cooling power. In contrast, the DSVFS system solves this problem, yielding daytime energy saving up to 200 $\mathrm{kWh/m^2}$. This cooling performance represents 70% of the fundamental limit achieved by the ideal emitter oriented horizontally (see supplemental Information, Section 7).

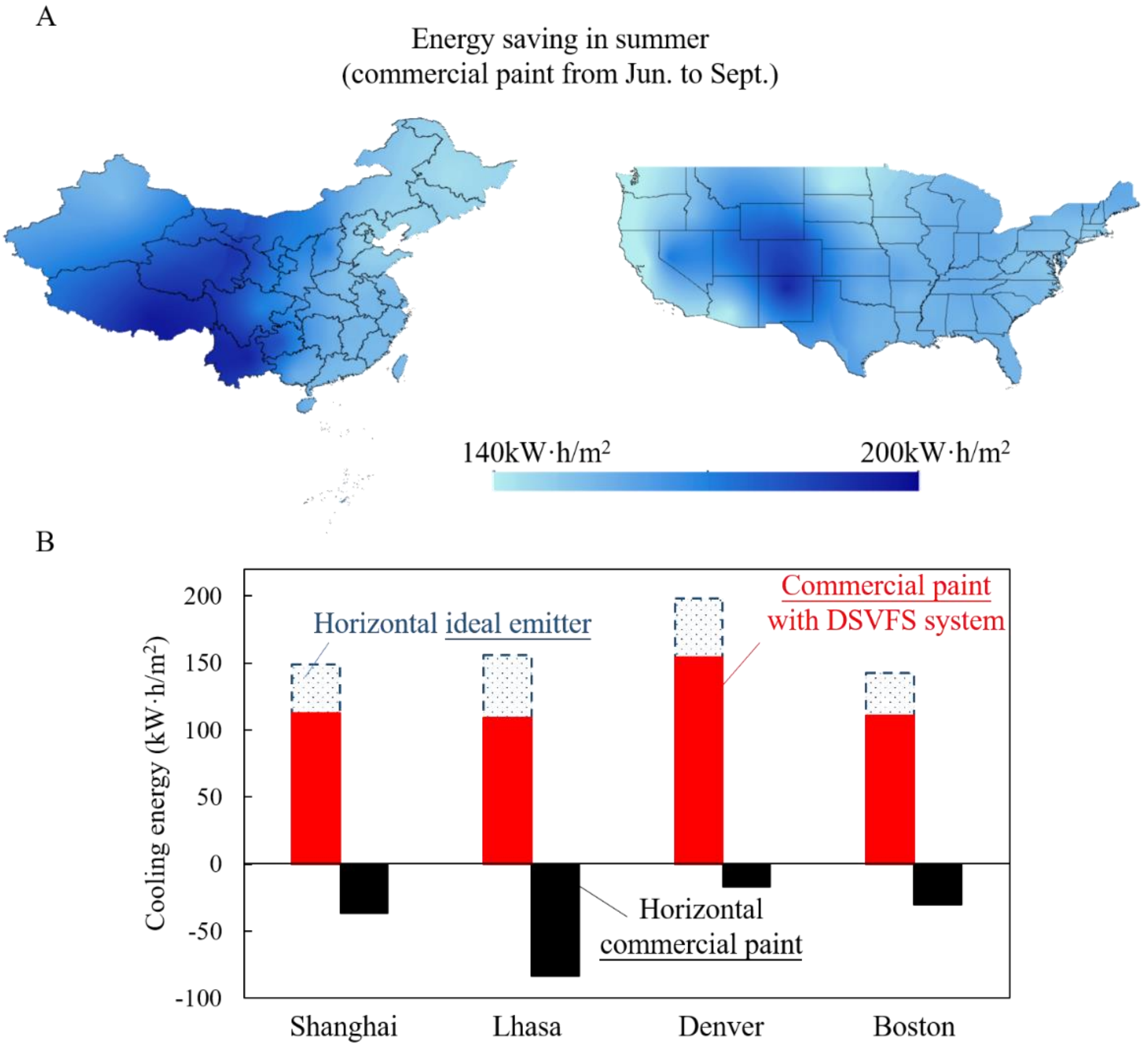

**Figure 5. Potential Global Cooling Energy Saving with DSVFS system.**

**(A)** Calculation samples cities across China (left) and the USA (right) in summer (from June to September). The color coding represents the difference between the cooling energy produced by a commercial paint (see figure S9 for its spectrum) equipped with DSVFS system and this same paint oriented horizontally: the darker the more energy savings.

**(B)** Four representative cities from China and USA to illustrate the cooling energy generated by an ideal emitter (blue dotted) with unity emissivity between 8 and 13 μm and zero emissivity (absorptivity) in other wavelengths oriented horizontally, the commercial paint with DSVFS system (red), and the commercial paint aligned horizontally (black). This comparison underscores the universal applicability of DSVFS system in practical application.

# Conclusions and Discussion

We experimentally achieved continuous 24-hour sub-ambient radiative cooling using a near-blackbody emitter (solar absorptivity $\alpha = 99.6\%$) equipped with the bio-inspired dynamic sky view factor steering (DSVFS) system. During a representative hot noon, this system achieved a mean temperature reduction of 5.1 °C below ambient and a mean cooling power of 36.9 W/m$^2$. In another experiment with a homemade selective radiative cooling paint with solar absorptivity of 4.6%, this DSVFS configuration led to a 135% enhancement in cooling flux at noon, as compared to its static horizontal counterpart. To demonstrate the large-scale potential of our approach, we use a commercial paint with solar absorptivity of 24.3% with the DSVFS system, yielding over 100 $\mathrm{kWh/m^2}$ of daytime cooling energy savings compared to the static horizontal counterpart across four representative cities during summer months (from June to September). This performance approaches to more than 70% of the fundamental limit that could be achieved by an ideal emitter oriented horizontally. Summer cooling energy saving exceeds 150 $\mathrm{kWh/m^2}$ in many regions of China and the USA, with a maximum up to 200 $\mathrm{kWh/m^2}$.

This DSVFS configuration establishes a simple, scalable, and low-cost pathway for efficient daytime radiative cooling, enabling spatiotemporal enhancement across all orientations, diurnal cycles, and weather conditions. Realizing a paradigm shift from static spectral optimization to dynamic spatiotemporal modulation, this DSVFS configuration allows broader material utilization for high-performance energy-saving cooling. Moreover, it overcomes a major limitation of traditional static configurations, which cannot reorient to maximize cooling performance in a 24-hour day–night cycle.

We end by commenting on the prospect for practical applications. First, although this DSVFS system requires additional electrical power to operate, but considering the fact that the net energy saving is at least an order of magnitude larger than the power input, the harvested cooling energy could potentially be used for the energy input. Moreover, integrating photovoltaic cells with this DSVFS configuration could enable simultaneous radiative cooling and solar power generation without external energy input[32]. Second, scaling up a single DSVFS module might be challenging, but this issue can be resolved by arranging multiple DSVFS units in arrays, which is very similar to the conventional tilted solar panel layout. Third, dynamic temperature regulation could be achieved by modulating the field-of-view orientation to maintain specific emitter temperatures, thereby enabling both solar heating and cooling.

**Data availability**

The data that support this research's findings are available and can be provided based on the request to the corresponding authors.

## Section 1: Materials and Methods

Radiative cooling paint

We fabricated radiative cooling paint (figure 3A in the main text) using a solution-processed coating method where polydimethylsiloxane (PDMS, Sylgard 184, Dow Inc.) served as the polymer matrix for dispersing zirconium dioxide microspheres ($ZrO_2$, Shanghai Yaoyi Alloy Material Co., Ltd.). The process initiated by mixing PDMS base and curing agent at a 10:1 weight ratio, followed by addition of $ZrO_2$ powder at 40 wt% solids loading and acetone. This mixture underwent mechanical stirring at 500 rpm for 90 minutes at 25°C to achieve uniform dispersion. The resulting precursor was doctor-bladed onto pre-cleaned aluminum substrates then thermally cured at 80°C for 2 hours under ambient atmosphere to form the final coating.

Other materials

Commercial paint was purchased from House Doctor (Shanghai) Technology Co., LTD. Near-blackbody paint was purchased from Foshan Shixiong Li New Material Technology Co., Ltd.

We employ nanoPE as an auxiliary covering to achieve blackbody daytime radiative cooling, yet using standard PE and double-layer nanoPE can also cool a blackbody emitter compared to the horizontal emitter (figure S1).

Optical characterization

Reflectivity spectra across the UV-Vis-NIR range (300-2500 nm) were characterized using a spectrophotometer (Agilent Cary 6000i) equipped with an internal integrating sphere (DRA-2500) to capture total reflectance, including both specular and diffuse components. Infrared absorptance measurements and total transmittance were obtained using Fourier-transform infrared spectroscopy (FTIR, Thermo Fisher is50) with a gold-coated integrating sphere attachment.

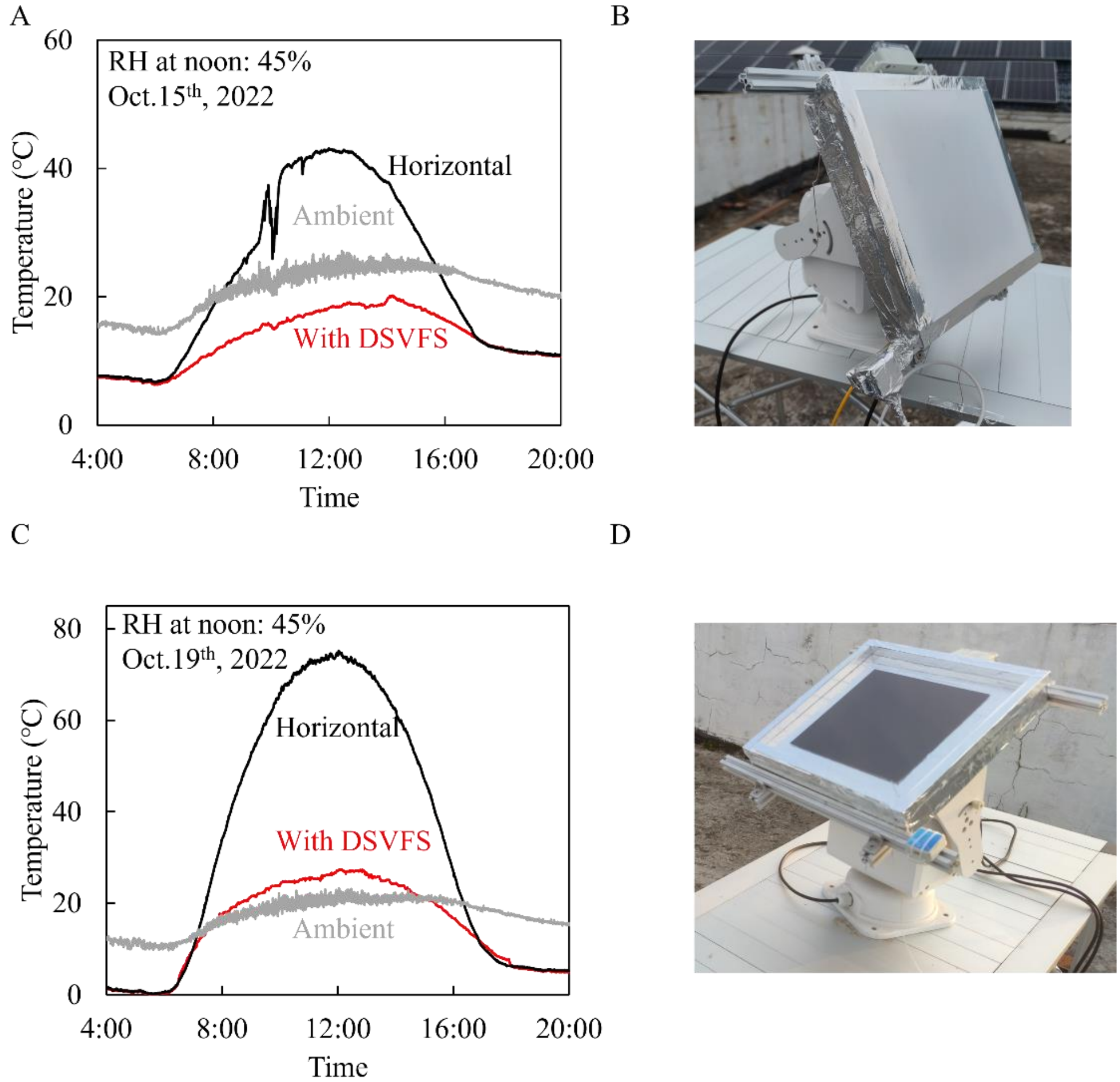

**Figure S1. Experiment results of black paint with double-layer nanoPE and regular PE**

(A) With double-layer nano-porous polyethylene (nano-PE) cover: subambient cooling achieved at solar noon.

(B) With regular PE cover: failure to maintain subambient cooling, motivating adoption of single-layer nano-PE as optimized diffused solar reflector.

## Section 2: Outdoor cooling tests

Emitters were simultaneously tested in both static-horizontal DSVFS and conventional horizontal configurations, with experimental uniformity validated through calibration (figure S2).

Temperature monitoring employed pre-calibrated K-type thermocouples connected to a data logger (AZ88598, AZ Instrument Corp; ±0.3% accuracy), while solar irradiance was measured with a pyranometer (SN-300AL-RA-N01, Prsens) and relative humidity recorded via a hygrothermograph (GSP-4, Elitech).

Cooling flux quantification (figure S3) involved measuring electrical power input to insulated flexible heaters bonded to aluminum emitters' rear surfaces. Each heater, powered by an 8-V DC supply and regulated by a PID controller (PT100, Anthone), maintained emitter temperature at ambient levels - enabling real-time cooling flux determination through heat input calculation.

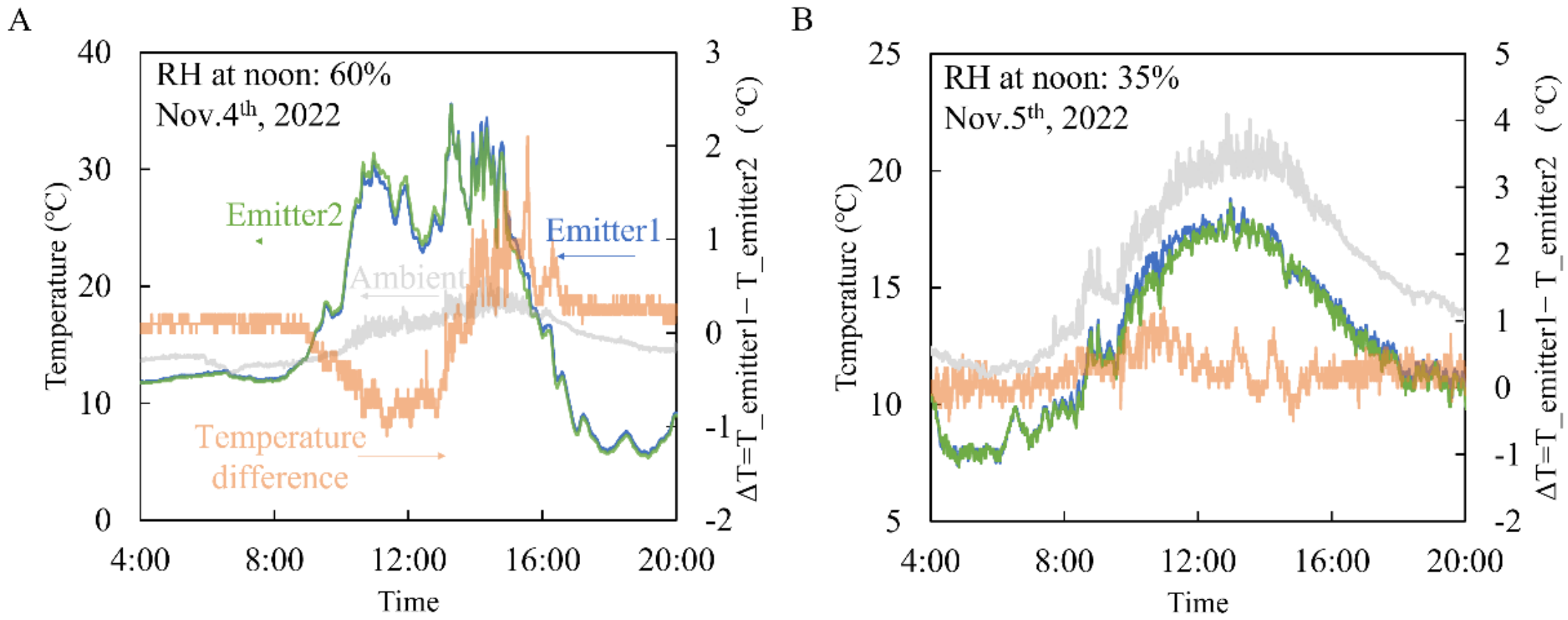

**Figure S2. Uniformity validation of devices maintained horizontally throughout diurnal testing.** Temperature profiles of near-blackbody emitter devices (A) and radiative cooling paint (B) demonstrate consistent performance between identical units, confirming experimental reliability. Device temperatures (green/blue), inter-device ΔT (orange), and ambient temperature (grey). Temperature profiles demonstrate mean inter-device ΔT <0.5°C, confirming experimental reliability.

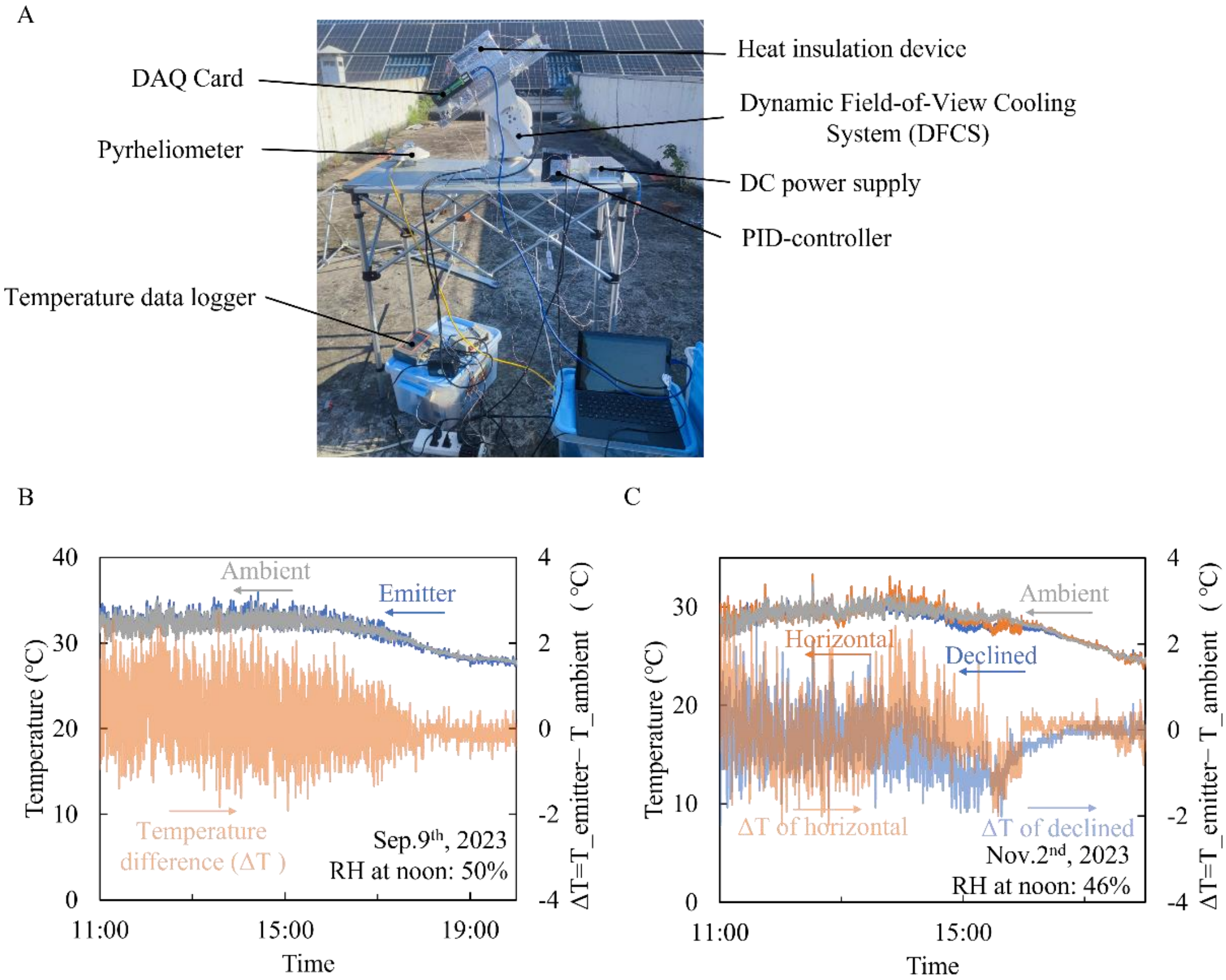

**Figure S3. Radiative cooling flux quantification**

(A) Experimental setup with PID controller maintaining emitter at ambient temperature.

(B) Near-blackbody emitter: Temperature profiles declined emitter (blue), ambient (grey), ΔT (orange).

(C) Radiative cooling paint: Temperature profiles (orange: declined, blue: horizontal), ambient (grey); ΔT (orange: declined, blue: horizontal). Maximum ΔT < 2℃ enabled accurate cooling flux measurement.

The daytime radiative cooling flux $P_{cooling}(T,\beta)$ of an emitter at temperature $T$ can be expressed by considering all contributions to the energy balance:

$$P_{cooling}(T,\beta) = P_{rad}(T) - P_{atm}(T_{amb},\beta) - P_{ground}(T_g,\beta) - P_{solar}(\beta) - P_{para}(T,T_{amb}) \quad \text{(S1)}$$

Here, $P_{rad}$ represents the radiated power by the emitter, $P_{atm}$ represents the absorbed atmospheric radiation by the emitter, $P_{ground}$ represents the absorbed ground radiation by the emitter, and $P_{solar}$ represents the absorbed solar irradiation by the emitter, $P_{para}$ represents parasitic heat exchange with the environment.

**Section 3: Solar radiance on declined surface**

$$P_{solar} = A\int_0^{\infty} \mathrm{d}\lambda \varepsilon(\lambda)\, I_{solar-declined}(\lambda) \quad \text{(S2)}$$

We use Perez model to calculate $I_{solar-declined}$, which can be segmented into direct beam and diffuse components, namely isotropic part, circumsolar diffuse and horizon brightening (figure S4). Equation S3-S11 describes the solar energy on declined surface.

$$I_{solar-declined} = I_b R_B + I_d(1-F_1)\left(\frac{1+\cos\beta}{2}\right) + I_d F_1 \frac{a}{b} + I_d F_2 \sin\beta + Ir_{ground}\left(\frac{1-\cos\beta}{2}\right) \quad \text{(S3)}$$

where $I_b$ represents the direct beam radiance on horizontal surface (Equation S3-6). $R_B$ represents the geometric factor, the ratio of beam radiation on the tilted surface to that on a horizontal surface at any time, equals $\frac{\cos\beta}{\cos\theta_z}$. $I_d$ represents the total diffuse components on horizontal surface (equation S4-7), Brightness coefficients $F_1$ and $F_2$ are functions of clearness $\varepsilon$ (equation S8-11). $a$ and $b$ are terms that account for the angles of incidence of the cone of circumsolar radiation, $a = \max(0, \cos\theta)$, $b = \max(\cos 85°, \cos\theta_z)$.

The decomposition of total solar radiation incident on a horizontal surface into its diffuse ($I_d$) and beam ($I_b$) components is conventionally achieved by correlating the diffuse fraction ($\frac{I_d}{I}$) with the clearness index ($k_T$). This empirical relationship leverages the fact that $k_T$, defined as the ratio of total horizontal solar radiation ($I$) to extraterrestrial horizontal radiation ($I_O$), serves as a robust indicator of atmospheric attenuation and cloud cover.

$$\frac{I_d}{I} = \begin{cases} 1 - 0.09k_T & for\ k_T \le 0.22 \\ 0.9511 - 0.1604k_T + 4.388{k_T}^2 - 16.638{k_T}^3 + 12.336{k_T}^4 & for\ 0.22 < k_T \le 0.8 \\ 0.165 & for\ k_T > 0.8 \end{cases} \quad \text{(S4)}$$

$$I_b = I - I_d \quad \text{(S5)}$$

where $k_T$ is clearness index,

$$k_T = \frac{I}{I_O} \quad \text{(S6)}$$

where $I$ is measurements of total solar radiation on a horizontal surface, which is commonly measured by pyranometer. $I_O$ represents extraterrestrial normal-incidence radiation.

$$I_O = G_{SC}(1.000110 + 0.034221\cos B + 0.001280\sin B$$
$$+0.000719\cos 2B + 0.000077\sin 2B) \quad (S7)$$

where $G_{SC}$ represents solar constant, here we take 1367 W/m$^2$, $B$ is given by $B = (n-1)\frac{360}{365}$.

The brightness coefficients $F_1$ and $F_2$ are parameters that describe the sky conditions.

$$F_1 = max\left[0, \left(f_{11} + f_{12}\Delta + \frac{\pi\theta_z}{180}f_{13}\right)\right] \quad (S8)$$

$$F_2 = \left(f_{21} + f_{22}\Delta + \frac{\pi\theta_z}{180}f_{23}\right) \quad (S9)$$

$f_{11}$, $f_{12}$, $f_{13}$, $f_{21}$, $f_{22}$, $f_{23}$ are brightness coefficients with recommend set (details see ref.# Table 2.16.1). $\Delta$ represents brightness parameter, $\Delta = \frac{I_d}{\cos\theta_z \cdot I_O}$, $\theta_z$ represents the zenith angle,

$$cos\theta_z = cos\phi cos\delta cos\omega + sin\phi sin\delta \quad (S10)$$

where $\phi$ represents the angular location north or south of the equator (north positive; $-90° \le \phi \le 90°$), $\omega$ represents the angular displacement of the sun east or west of the local meridian due to rotation of the earth on its axis at $15°$ per hour (morning negative, afternoon positive), $\delta$ represents the angular position of the sun at solar noon with respect to the plane of the equator (north positive; $-23.45° \le \delta \le 23.45°$), $\delta = 23.45 \sin\left(360\frac{284+n}{365}\right)$, $n$ represents the $n$th day of the year.

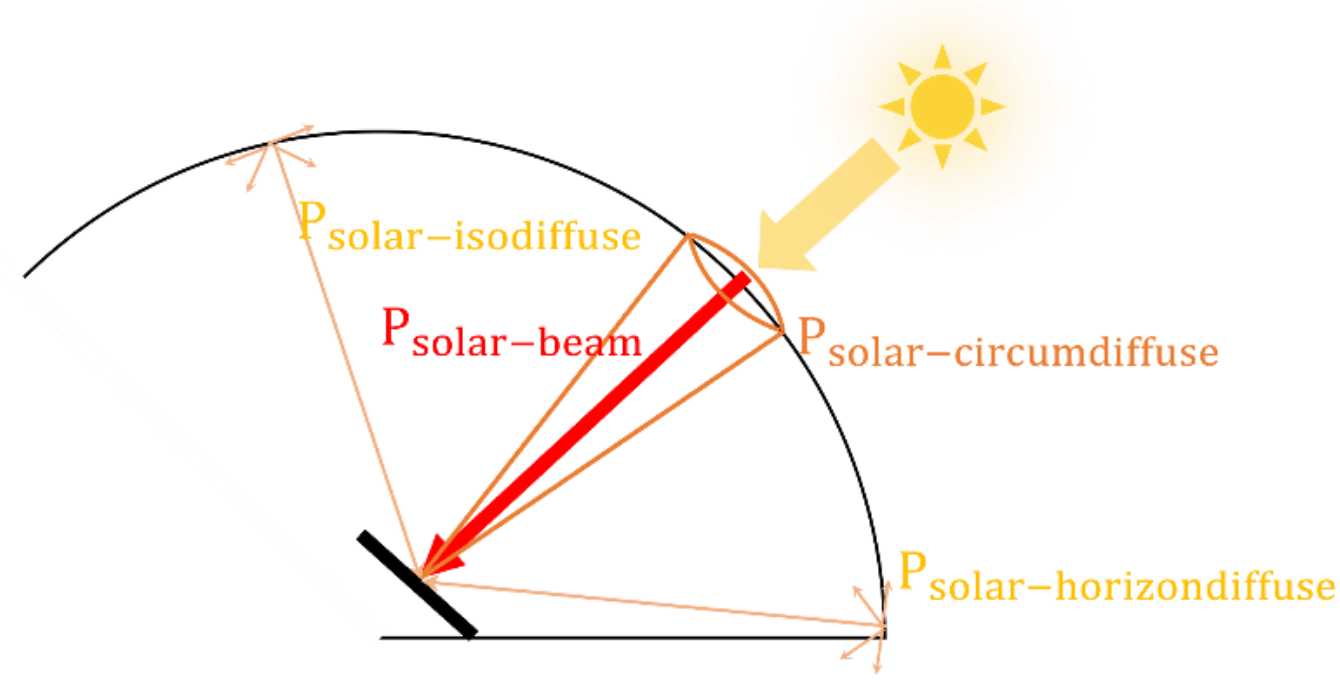

**Figure S4. Solar Radiance Model on Declined Surface**

The total radiation on the declined surface includes five terms: the beam, the isotropic diffuse, the circumsolar diffuse, the diffuse from the horizon, and the ground-reflected term.

**Section 4: Environmental radiance on declined surface**

Figure S5A shows the radiation heat transfer with the environment. The thermal emission from the declined emitter above the horizon exchanges heat with the atmosphere while that below the horizon exchanges heat with the ground.

As illustrated in figure S5A, the integration domain for atmospheric heat exchange comprises a radiation hemisphere truncated by the horizon plane (red solid line). We partition this hemisphere into two sectors using the x-z plane. The sector extending along the positive y-axis constitutes a complete spherical quadrant (full quarter-sphere), while the sector along the negative y-axis forms a spherical quadrant bounded by both the horizon plane (green) and the x-z plane. The atmospheric radiance absorbed by the emitter can therefore be evaluated according to equation S11.

$$p_{atm}(T_{amb},\lambda,\theta,\varphi)=\int_0^{\pi}d\varphi\int_0^{\theta(\varphi)}I_{atm}(T_{amb},\lambda,\gamma,H)\varepsilon_{emitter}(\lambda,\theta)\sin\theta\cos\theta\,d\theta d\lambda$$
$$+\int_{\pi}^{2\pi}d\varphi\int_0^{\frac{\pi}{2}}I_{atm}(T_{amb},\lambda,\gamma,H)\varepsilon_{emitter}(\lambda,\theta)\sin\theta\cos\theta\,d\theta d\lambda \quad \text{(S11)}$$

where $\theta(\varphi)$ represents the function of the declined surface ($\sin^2\theta=\frac{1}{1+\sin^2\varphi\tan^2\beta}$). $I_{atm}$ represents the directional atmospheric radiance, which could be calculated by Modtran[28]. H represents the altitude. In the coordinate system aligned with the declined surface. $\gamma$ represents the angle of atmospheric in the horizontal coordinate ($x^{'}, y^{'}, z^{'}$).

To evaluate radiative heat transfer through integration from the emitter's perspective, we require the zenith angle ($\gamma$) of atmospheric radiance in the horizontal coordinate system (see figure S5B). $\gamma$ is obtained by computing the angle between the normal vector of the declined surface ($\overrightarrow{n_1}=\{0,-\tan\beta,1\}$) and the radial direction vector ($\overrightarrow{n_2}=\{sin\,\theta\,cos\,\varphi, sin\,\theta\,sin\,\varphi, cos\,\theta\}$) normal to the spherical surface.

$$\cos\gamma=\cos\langle\overrightarrow{n_1}|\overrightarrow{n_2}\rangle=\frac{-\sin\theta\sin\varphi\tan\beta+\cos\theta}{\sqrt{1+\tan^2\beta}} \quad \text{(S12)}$$

Similarly, the integration domain for ground heat exchange comprises the lower radiation hemisphere truncated by the horizon plane, extending beneath it to the ground surface (red dashed line).

$$p_{ground}(T_{ground},\lambda,\theta,\varphi)=\int_0^{\pi}d\varphi\int_{\theta(\varphi)}^{\frac{\pi}{2}}I_{BB}(T_{ground},\lambda)\varepsilon_{ground}\varepsilon_{emitter}(\lambda,\theta)\,sin\,\theta\,cos\,\theta\,d\theta d\lambda \quad \text{(S13)}$$

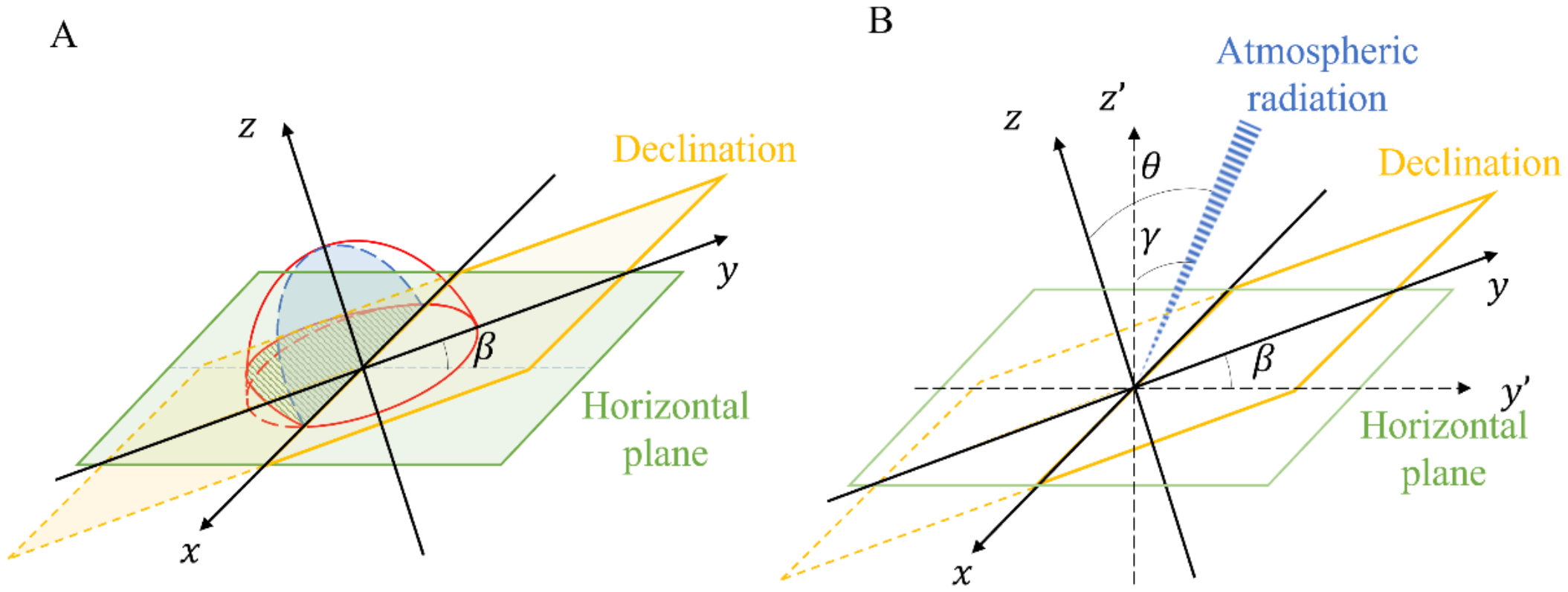

**Figure S5. Environmental Radiance Model on Declined Surface**

(A) The environmental radiance integration domain for atmospheric heat exchange covers the radiation hemisphere truncated by the horizon plane (solid red line), whereas that for ground heat exchange encompasses the lower radiation hemisphere truncated by the horizon plane and extending beneath it to the ground surface (dashed red line).

(B) For coordinate transformation of angles, γ represents the atmospheric radiance angle in the horizontal coordinate system (x', y', z'). This angle must be derived from the atmospheric radiance zenith angle (θ) and the declination angle (β) to calculate radiative heat transfer from the emitter's perspective.

**Section 5: Comparison between experiments and theory**

To validate the theoretical model's accuracy, we compare its predictions with experimental measurements by inputting test data—including emitter spectral properties, solar irradiance, relative humidity, surface tilt angle, and precise spatiotemporal coordinates (time/location)—then calculating the resulting emitter temperature and cooling flux.

The first row of figure S6 summarizes the $\Delta T$ as a function of horizontal solar intensity in field experiments in Nanjing on Nov. 2nd, 2022 and Mar. 15th, 2023. The x-axis error bar (5%) was estimated based on pyrheliometer and the y-axis error bar ($\pm 1$°C) is from the datasheet of the K-type thermocouple. The experimental results which picked randomly (marked circle) are consistent in magnitude our with theoretical model (The shaded region represents the wind-dependent variation in convective heat transfer coefficient h, with value ranging from 4.9 $Wm^{-2}K^{-1}$ to 12.3 $Wm^{-2}K^{-1}$ for blackbody paint and from 7.6 $Wm^{-2}K^{-1}$ to 15.1 $Wm^{-2}K^{-1}$ for radiative cooling paint).

The second row of figure S6 exhibits the cooling power as a function of horizontal solar intensity in field experiments in Nanjing on Sep. 9th, 2022 and Nov. 2nd, 2023. The y-axis error bar is the error of temperature difference between the emitter and the ambient caused by the thermocouple which affected the precise of PID-controller and the final cooling power. The experimental results which picked randomly (marked circle) are consistent in magnitude our with theoretical model. (Shaded area where error of ambient temperature, ground temperature and humidity are taken into consideration)

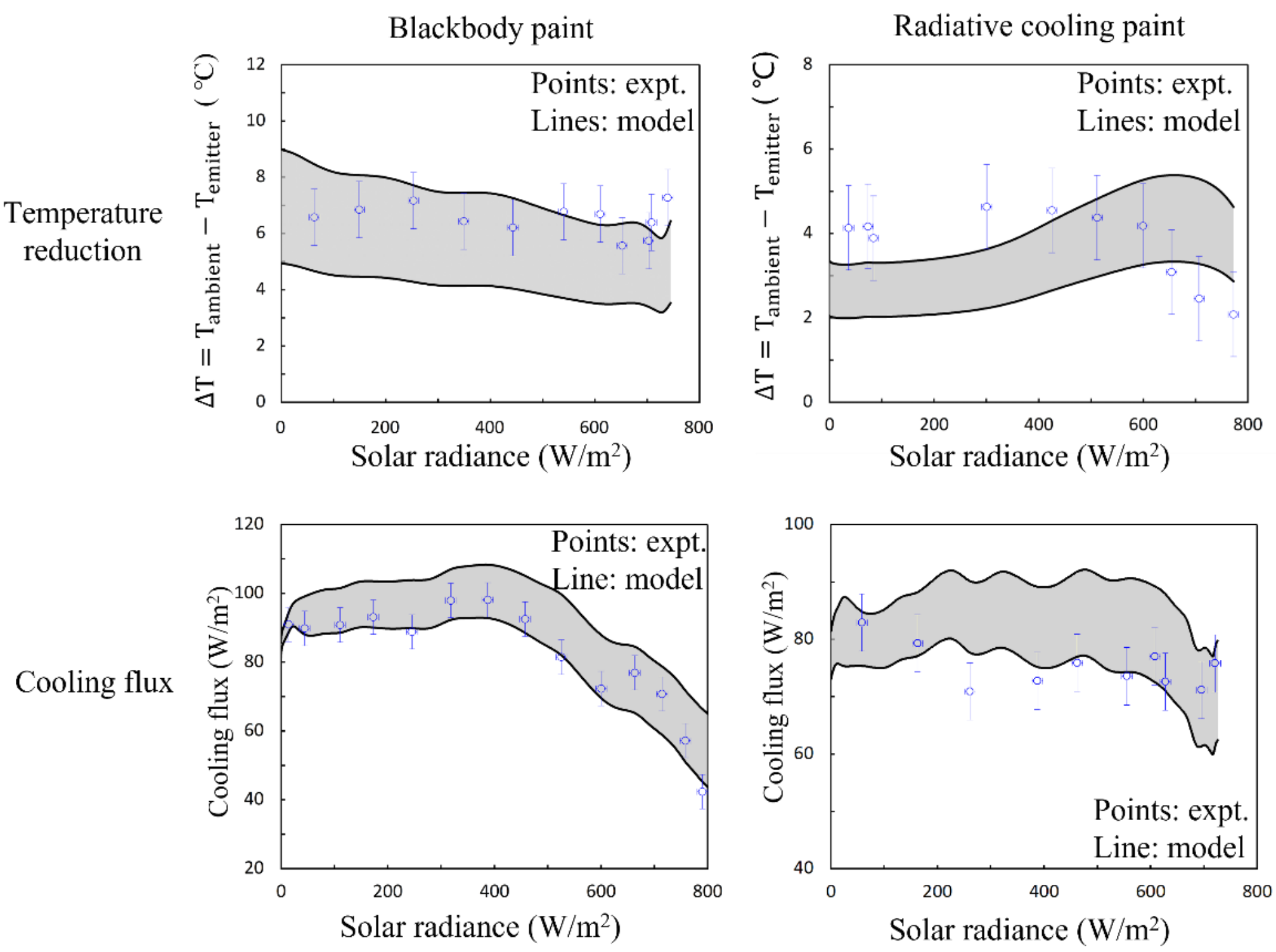

**Figure S6. Comparison between experiments and theory.**

The first row shows the temperature reduction comparison as a function of horizontal solar intensity in field experiments in Nanjing on Nov. 2nd, 2022 and Mar. 15th, 2023. The second row shows the cooling power comparison as a function of horizontal solar radiance in field experiments in Nanjing on Sep. 9th, 2022 and Nov. 2nd, 2023.

**Section 6: Relationship between optimal declined angle and spectral selectivity**

Figure S7 shows the cooling flux of different emitters under different solar zenith angle with DSVFS.

Emitters with emittance of 0.95 in 8-13$\mu m$ and reflectivity of 0.95, 0.96, 0.97, 0.98, 0.99 and 1 in 0.3-4μm are selected to calculate the cooling power variation under solar zenith angle of 10°, 30°, 50°, 70°.

As is shown in figure S7A, for all the emitters except ideal case ($R_{0-4\mu m} = 1$), cooling flux increases with the declined angle at the beginning. This occurs because solar shielding gains outweigh infrared sky-view losses. The dashed black line indicates each emitter's maximum cooling flux at its optimal declined angle. The optimal declined angle increases as solar reflectivity decreases, approaching the solar elevation angle at the limit ($\beta = \eta = 90° - \theta_z$).

Figures S7B illustrates the relationship between the optimal declined angle and solar elevation angle. For emitters with solar reflectivity smaller than 0.96 ($R_{0-4\mu m} < 0.96$), the optimal angle consistently equals the solar elevation angle ($\beta = \eta = 90° - \theta_z$) as shown by black line. As solar elevation increases, the critical reflectivity threshold (where $\beta < \eta$) decreases. This reveals a counterintuitive relationship between solar elevation and optimal emitter orientation. At lower solar elevation angles, high-reflectivity emitters must align nearly parallel to incident sunlight to maximize cooling performance. However, as solar elevation increases and direct irradiance intensifies, emitters progressively revert toward horizontal alignment.

Figure S7C elucidates this phenomenon by demonstrating how absorbed environmental radiance (black) increases while solar irradiance (grey, parameterized by solar zenith angle $\theta_z$) decreases relative to a horizontal surface as emitter declined angle increases, shown here for an emitter with $R_{0-4\mu m} = 0.98$. Since the emitted radiative cooling flux and parasitic heat remain decline-angle-invariant, the point of maximum differential between environmental radiance (black) and solar irradiance (grey) corresponds to the optimal cooling flux. Crucially, environmental radiance variation with declination is independent of solar position. However, for smaller zenith angles (higher solar elevations), the initial reduction in solar absorption at low declined angles is less pronounced relative to horizontal. Consequently, high-reflectivity emitters exhibit stronger preference for horizontal alignment under smaller zenith angle conditions.

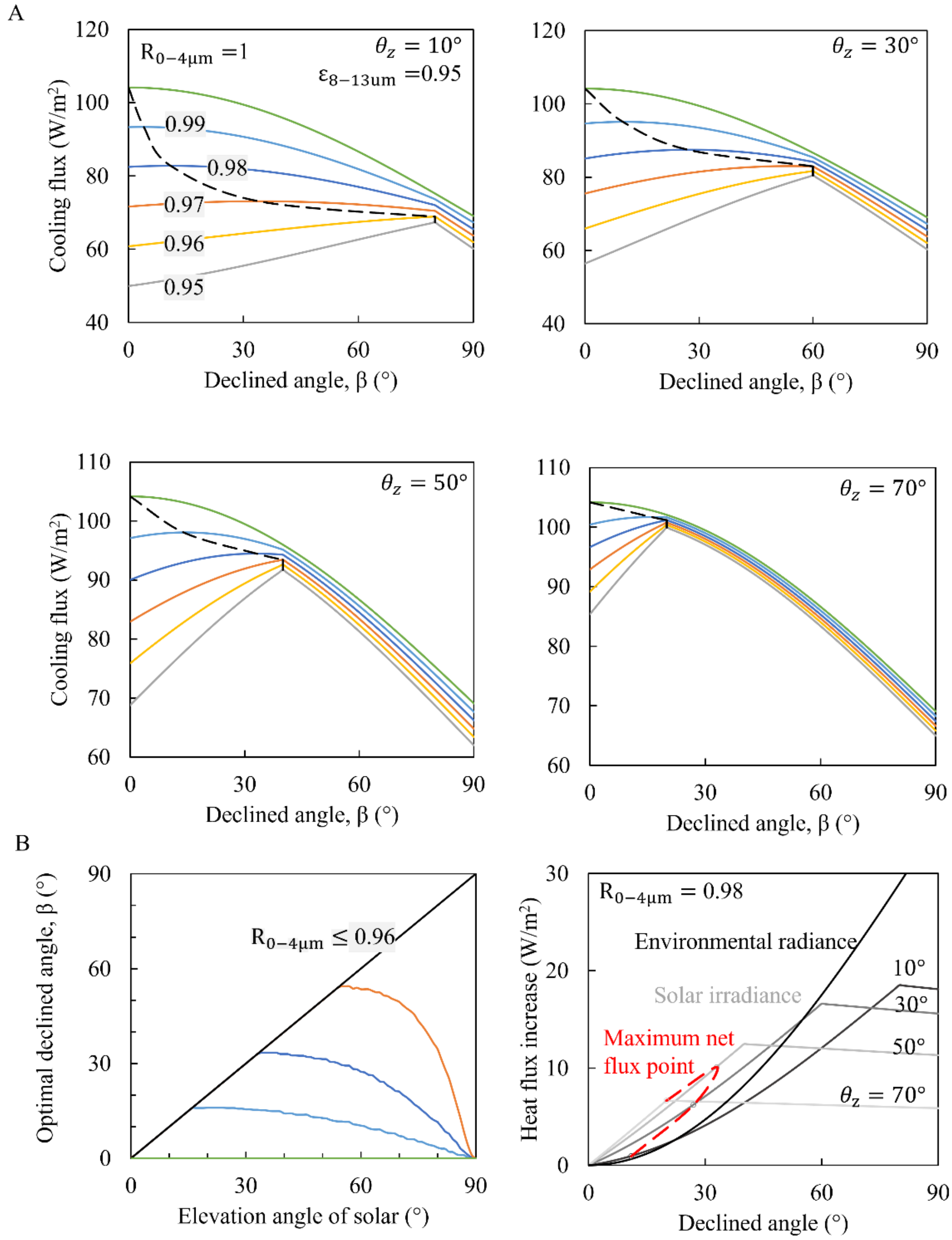

**Figure S7 Relationship between optimal declined angle and spectral selectivity**

(A) Cooling flux as a function of emitter declination angle for varying solar reflectivity ($R_{0-4\mu m}$ = 1, 0.99, 0.98, 0.97, 0.96, 0.95) and solar zenith angles ($\theta_z$ = 10°, 30°, 50°, 70°). The dashed grey curve denotes the maximum cooling flux across all configurations.

(B) Optimal declination angle for maximum cooling flux across emitters with different $R_{0-4\mu m}$, plotted against solar elevation angle (90° - $\theta_z$).

(C) Trade-off analysis: increasing absorbed environmental radiance (black) and reducing solar irradiance (grey, parameterized by $\theta_z$), normalized to a horizontal surface, versus increasing declined angle. Net flux maximization occurs at the largest difference between these components (red).

**Section 7: Daily accumulated cooling energy and summer daytime energy increment**

Applying our declined-surface heat transfer model, we compute real-time radiative cooling flux for Nanjing using nanoPE-coated black paint emitters under two configurations: with and without a Dynamic Sky View Factor System (DSVFS). To quantify annual enhancement potential, figure 2D presents integrated daily cooling energies for four seasonally critical dates—March 23 (vernal equinox), June 23 (summer solstice), September 23 (autumnal equinox), and December 23 (winter solstice)—contrasting DSVFS-optimized performance against the static-horizontal configuration.

To verify the universality of location and real cooling effects of DSVFS, we choose 31 provincial capitals in China and 50 state capitals in the US and calculate the cooling energy enhancement with DSVFS during summer time (from June to September) with commercially available paint as emitter. The energy enhancement was represented by using accumulated cooling flux with DSVFS during the period minus that of horizontal. The cooling flux of both are equal at night so the enhancement amount is that during day time. The location influences the cooling flux amount though longitude, latitude, altitude, temperature and humidity (data was obtained from https://www.weather-atlas.com). Figure 5A is plotted by interpolate the above data points using the inverse distance weighting method in the range of China and the US.

To verify the geographical universality and practical efficacy of the Dynamic Sky View Factor System (DSVFS), we evaluated cooling energy enhancements across 31 Chinese provincial capitals and 50 U.S. state capitals during summer (June–September). Using commercially available paint emitters, we quantified enhancement as: $\Delta q = q_{with\ \mathrm{DSVFS}} - q_{static-horizontal}$. Location-specific parameters—longitude, latitude, altitude, temperature, and humidity (sourced from https://www.weather-atlas.com)—were incorporated to model regional cooling flux variations. Figure 5A spatially interpolates these results across China and the U.S. using Inverse Distance Weighting (IDW) methodology:

$$P(u) = \frac{\sum_{i=1}^{n} \frac{P_i}{d_i^2}}{\sum_{i=1}^{n} \frac{1}{d_i^2}} \tag{S14}$$

where $P(u)$ is the cooling flux of the interpolation point u, $P_i$ is cooling flux enhancement at the i-th capital city, $d_i$ is the distance between interpolation point u to the i-th known location, n is the total number of reference points (31 for China and 50 for the US)

Applying this methodology, we selected four representative cities—Shanghai and Boston (eastern coastal, low-altitude) versus Lhasa and Denver (western interior, high-altitude)—to quantify DSVFS performance across geographic extremes. Cooling energy enhancements were quantified against horizontal commercial paints and ideal emitters characterized by solar reflectivity $\rho$=1 (0.3-4 μm) and thermal emissivity $\varepsilon$=1 (4-30 μm).

Critical to our analysis was the implementation of MODTRAN 6 for calculating infrared downwelling irradiance, which supersedes transmittance-based cosine approximations known to underestimate cooling potential by >10%. This advanced radiative transfer model accounts for atmospheric thermal inhomogeneity, revealing significant net cooling flux contributions beyond the 8-13 μm atmospheric window when emitter temperature equals ambient. Consequently, our definition of the ideal emitter ($\varepsilon$=1 across 4-30 μm)) captures the full-spectrum radiative cooling potential, exceeding traditional atmospheric window-selective emitters with greater cooling flux.

## Section 8: Daily accumulated cooling energy and summer daytime energy increment

Unlike dynamic solar shields where area scaling induces obstruction of the sky view causing radiative energy degradation, the DSVFS configuration enables scalability without energy loss through modular array deployment (Figure S8).

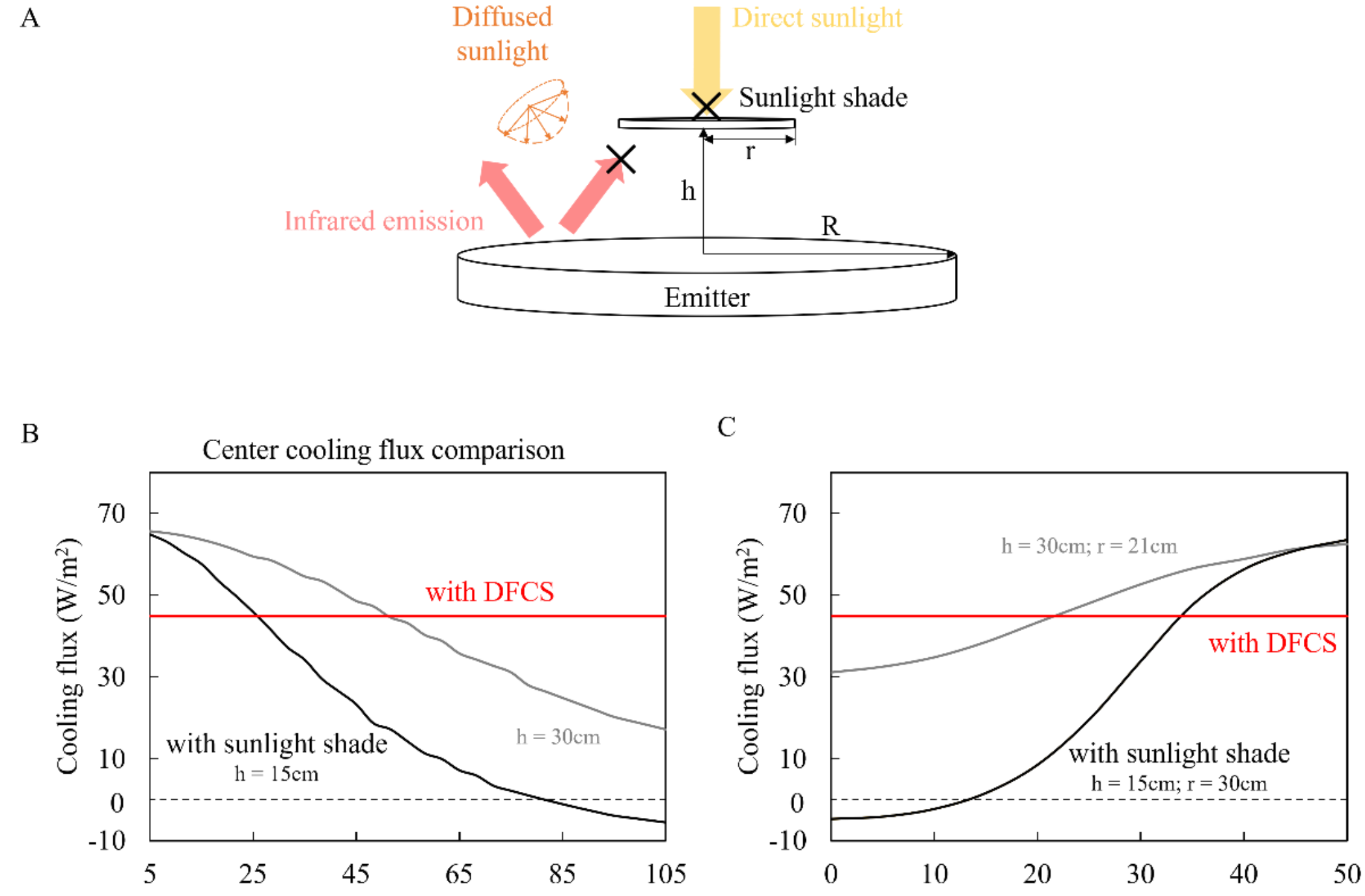

**Figure S8. Cooling flux comparison: Solar shade vs. DFCS in large area**

(A) Schematic of shade-cooling mechanism blocking direct sunlight while partially obstructing infrared skyward emission.

(B) Shaded emitter shows center flux reduction (infrared blockage amplified with scaling) vs. DFCS maintaining constant flux.

(C) At 15 cm shade height, 50 cm emitter requires 30 cm shade radius, exhibiting strong center-to-edge flux nonuniformity.

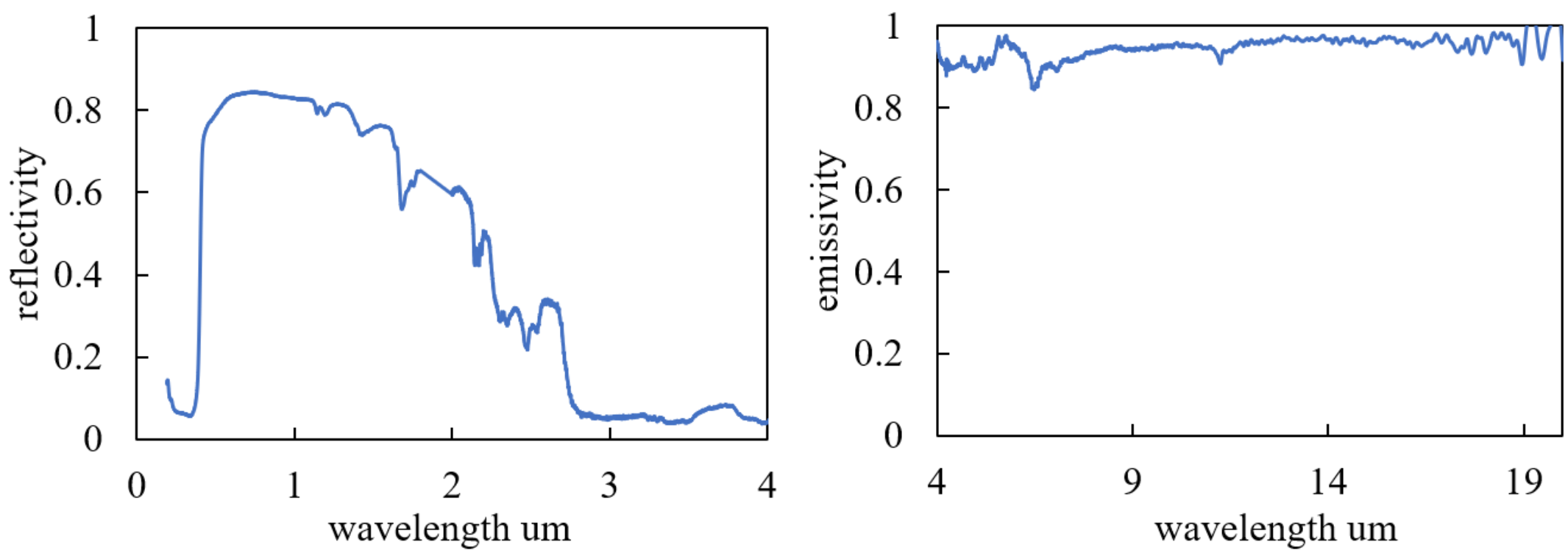

Fig S9. Spectrum of the commercial paint

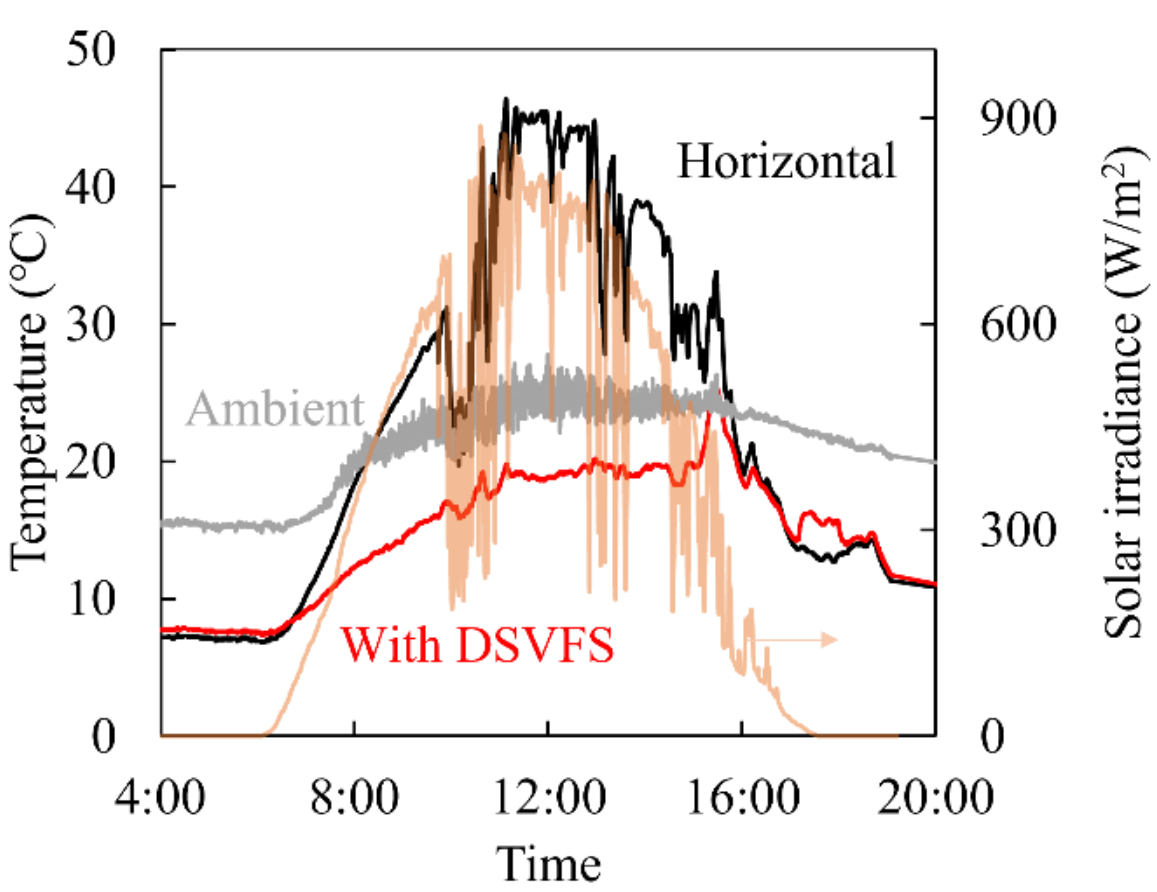

Fig S10. Field experimental under a cloudy day with blackbody emitter covered by nanoPE